\documentclass[fleqn,usenatbib]{rasti}

\usepackage{newtxtext,newtxmath}
\usepackage{float}
\usepackage{booktabs}
\usepackage{multirow}
\usepackage{longtable}
\usepackage[T1]{fontenc}
\usepackage[version=4]{mhchem}
\usepackage{array}
\usepackage{geometry}
\usepackage{ragged2e}
\usepackage{hyperref}
\usepackage{makecell} 
 \usepackage{color}
\usepackage{amsmath} % for math formatting
\usepackage{booktabs}
\usepackage{comment}
\usepackage{graphicx}
\usepackage{tabularx}
\usepackage{listings}
\DeclareRobustCommand{\VAN}[3]{#2}
\let\VANthebibliography\thebibliography
\def\thebibliography{\DeclareRobustCommand{\VAN}[3]{##3}\VANthebibliography}
\definecolor{pink}{RGB}{255, 105, 180} % Define a pink color
\usepackage{graphicx}	% Including figure files
\usepackage{amsmath}	% Advanced maths commands
\usepackage{color}

\lstdefinelanguage{json}{
    basicstyle=\ttfamily\small,       
    numbers=left,                     
    numberstyle=\tiny\color{gray},    
    stepnumber=1,                     
    numbersep=5pt,                    
    showstringspaces=false,           
    breaklines=true,                  
    frame=single,                     
    backgroundcolor=\color{gray!10},  
    literate=
     *{0}{{{\color{blue}0}}}{1}       
      {1}{{{\color{blue}1}}}{1}
      {2}{{{\color{blue}2}}}{1}
      {3}{{{\color{blue}3}}}{1}
      {4}{{{\color{blue}4}}}{1}
      {5}{{{\color{blue}5}}}{1}
      {6}{{{\color{blue}6}}}{1}
      {7}{{{\color{blue}7}}}{1}
      {8}{{{\color{blue}8}}}{1}
      {9}{{{\color{blue}9}}}{1}
      {:}{{{\color{red}{:}}}}{1}      
      {,}{{{\color{red}{,}}}}{1}      
       {"}{{{\color{green!50!black}{"}}}}{1},     
}

\title[The \textsc{ExoAtom} Database]{\textsc{ExoAtom}: A Database of Atomic Spectra in ExoMol Format}

\author[Q.-H. Ni et al.]{
Qing-He Ni,$^{1}$
Rujia Wang,$^{1}$
Tianyang Xie,$^{1}$
Jingxin Zhang,$^{1, 3}$
Christian Hill,$^{2}$
Sergei N. Yurchenko,$^{1}$
\newauthor{
and Jonathan Tennyson$^{1}$ \thanks{E-mail: j.tennyson@ucl.ac.uk}}
\\
$^{1}$Department of Physics and Astronomy, University College London, Gower Street, London WC1E 6BT, UK\\
$^{2}$International Atomic Energy Agency, Vienna A-1400, Austria\\
$^{3}$State Key Laboratory of High Temperature Gas Dynamics, Institute of Mechanics, Chinese Academy of Sciences, Beijing 100190, China\\
}

\date{Accepted XXX. Received YYY; in original form ZZZ}

\pubyear{\the\year{}}

\begin{document}
\label{firstpage}
\pagerange{\pageref{firstpage}--\pageref{lastpage}}
\maketitle

\begin{abstract}
We present the \textsc{ExoAtom} database, www.exomol.com/exoatom, an extension of the ExoMol database to provide atomic line lists in the ExoMol format. \textsc{ExoAtom} is designed for detailed astrophysical, planetary, and laboratory applications.
\textsc{ExoAtom} currently includes atomic data for 80 neutral atoms and 74 singly charged ions. These data are extracted from both the \textsc{NIST} and Kurucz databases, with 79/71 atoms/ions sourced from \textsc{NIST} and 38/37 atoms/ions sourced from Kurucz. \textsc{ExoAtom} uses the file types \texttt{.all}, \texttt{.def}, \texttt{.states}, \texttt{.trans} and \texttt{.pf} as fundamental components for structuring atomic data in a consistent hierarchy. The \texttt{.states} file contains quantum numbers, uncertainties, lifetimes, etc. The \texttt{.trans} file specifies  Einstein \(A\) coefficients and their associated wavenumbers. The \texttt{.pf} file provides partition functions over a wide grid of temperatures. Post-processing of the \textsc{ExoAtom} data is provided by the program \textsc{PyExoCross}.
Future development of \textsc{ExoAtom} will include additional ionization stages.

\end{abstract}

% Include between one and six keywords.

%%%%%%%%%%%%%%%%%%%%%%%%%%%%%%%%%%%%%%%%%%%%%%%%%%

%%%%%%%%%%%%%%%%% BODY OF PAPER %%%%%%%%%%%%%%%%%%
\onecolumn

\section{Introduction}

The ExoMol database was established in 2011 to provide molecular line lists for exoplanet and other atmospheres
\citep{jt528}. Given their importance the ExoMol database  included contributions due to atomic  sodium \citep{19AlSpLe.atom} and potassium \citep{16AlSpKi.atom} in its opacity tables \citep{jt801}. However,
spectra of many neutral and singly ionised atoms are  a regular feature of high resolution studies
of exoplanetary atmospheres  \citep{10LiYaFr,10FoHaFr,18SpSiEv.He,18HoEhKi,20HoSePi,24HoKiMo,23JiWaZh,23PeBeAl,24DaSaBo,24SiPaWa}. For example, \citet{25PrSeHo} considered the spectra of 89 neutral or singly ionised atoms
 in their analysis of the ultrahot Jupiter exoplanet WASP-121 b
 and assigned features due to 17 of them in a study in which they also searched for the spectral signature of TiO.
 The increasing need for atomic spectra to be used alongside molecular spectra in exoplanets, and indeed
 other astronomical objects, has led to users requesting us to expand the ExoMol database to explicitly include atomic line spectra. As this expansion more than
 doubles the number of species included in the ExoMol database and because there are some subtle differences in
 the data presented for atoms and molecules, we have chosen to create a new database section of the ExoMol database presenting atomic data called \textsc{ExoAtom}, while retaining the original ExoMol branding for the molecular
 line lists \citep{jt939}. As detailed below, \textsc{ExoAtom} uses the same data structure as the molecular section of the ExoMol database and
 retains many of its characteristics.

 There are a number of existing databases which provide atomic spectroscopic data which are used in astronomical studies.
 These include the Atomic Spectra Database of the National Institute of Standards and Technology \citep{ralchenko2020development} (henceforth simply \textsc{NIST}), the line list compilations due to 
    \citet{kurucz2011including,Kurucz2017}, VALD3 (the third edition of the Vienna Atomic Line Database) \citep{VALD3},
 CHIANTI - An Atomic Database for Emission Lines   \citep{CHIANTI}
 and the Opacity Project \citep{05Seaton.OP} including the
 more focused spin-off Iron Project \citep{TIPTOPbase}. \textsc{ExoAtom} contains data extracted from the \textsc{NIST}
  and Kurucz databases. These two databases provided complementary information with \textsc{NIST} providing a more limited, high accuracy set of spectroscopic parameters largely based on precision laboratory measurements while the Kurucz line lists aim at completeness with many of the results coming from quantum mechanical calculations. 

Here we present the \textsc{ExoAtom} database, which aims  to expand the ExoMol database by providing atomic line lists  derived from both the \textsc{NIST} and Kurucz databases in the ExoMol format. 

\section{Data Coverage}
Below is a detailed description of the \textsc{ExoAtom} data coverage. Table \ref{amount} summarizes the final contents of the \textsc{ExoAtom} database. Table \ref{tb14} and Table \ref{tb15} list the total number of energy states and transition lines for each atom and ion, respectively, in the \textsc{NIST} dataset. Table \ref{tb6} and Table \ref{tb7} provide the corresponding counts of states and lines for each atom and ion in the Kurucz database.

\begin{table}
  \centering
  \caption{Data coverage of the \textsc{ExoAtom} database}
  \label{amount}
  \begin{tabular}{lrrrr}
    \hline
    Category             & Atom Count  & \textsc{NIST} & Kurucz \\
    \hline
    Neutral Atoms        & 80                     & 79           & 38              \\
    Singly Charged Ions  & 74                     & 71           & 37              \\
    \hline
  \end{tabular}
\end{table}

\begin{longtable}{lrrl}
    \caption{\label{tb14} Summary of neutral atoms (including isotopes) with \texttt{.trans} and \texttt{.states} files from \textsc{NIST} including numbers of states ($N_{\text{state}}$), number of transitions ($N_{\text{trans}}$) and source references.} \\
    \toprule
    Element & $N_{\text{state}}$ &  $N_{\text{trans}}$ & References \\
    \midrule
    \endfirsthead

    \multicolumn{4}{c}%
    {{\bfseries \tablename\ \thetable{} -- continued from previous page}} \\
    \toprule
    Element & $N_{\text{state}}$&  $N_{\text{trans}}$ & References \\
    \midrule
    \endhead

    \midrule \multicolumn{4}{r}{{Continued on next page}} \\
    \endfoot

    \bottomrule
    \endlastfoot

   $^{1}$H I & 105 & 441 & \cite{KRAMIDA2010586}, \cite{10.1063/1.1796671} \\
    $^{2}$H I & 77 & 161& \cite{KRAMIDA2010586}, \cite{8610TP} \\
    $^{3}$H I & 9&11 &  \cite{KRAMIDA2010586}, \cite{8610TP} \\
    $^{4}$He I & 842 &  2289 & \cite{PhysRevLett.105.063001}, \cite{8610TP} \\
    Li I & 181 & 257 & \cite{1987JPCRD..16S....K}, \cite{8610TP} \\
    Be I & 212 & 394 & \cite{PhysRevA.101.042503}, \cite{8702TP} \\
    B I & 118 & 242 &  \cite{Kramida_2007}, \cite{8702TP} \\
    C I & 433 & 1616 &  \cite{9787TP}, \cite{8028TP} \\
    N I &  366&1287  &  \cite{PhysRevA.109.052822}, \cite{8028TP} \\
    O I & 580 & 854 & \cite{6492TP}, \cite{DEMIRDAK2023111893} \\
    F I & 302 & 120 & \cite{PhysRevA.109.052822}, \cite{7043TP}  \\
    Ne I & 374 & 533 &  \cite{6976TP}, \cite{10.1063/1.1797771} \\
    Na I & 423 & 523 &  \cite{10.1063/1.2943652}, \cite{8098TP} \\
    Mg I & 303 & 1090 & \cite{8098TP}, \cite{10.1063/1.555617} \\
    Al I & 185 & 293 &  \cite{8098TP}, \cite{10.1063/1.555608} \\
    Si I & 406 & 603 & \cite{8363TP}, \cite{10.1063/1.555685} \\
    P I & 284 & 132 & \cite{1619TP}, \cite{10.1063/1.555736}  \\
    S I &377  &913  &  \cite{8595TP}, \cite{10.1063/1.555862} \\
    Cl I & 366 & 99 & \cite{1619TP}, \cite{Radziemski:69} \\
    Ar I & 503 & 428 &  \cite{6604TP},\cite{Minnhagen:73} \\
    K I & 296 & 211 & \cite{1619TP}, \cite{8198TP} \\
    Ca I & 780 & 136 & \cite{1619TP}, \cite{osti_6008300} \\
    Sc I & 355 &257  & \cite{5297TP}, \cite{osti_6008300}  \\
    Ti I & 558 & 496 &  \cite{5297TP}, \cite{10.1063/1.3656882} \\
    V I & 548 & 1162 &  \cite{9771TP} \\
    Cr I & 618 & 522 &  \cite{5297TP}, \cite{10.1063/1.4754694} \\
    Mn I & 536 & 489 &  \cite{5297TP}, \cite{osti_6008300} \\
    Fe I & 837 & 2542 &  \cite{7958TP}, \cite{1994ApJS...94..221N} \\
    Co I & 277 & 336 &  \cite{5298TP}, \cite{osti_6008300} \\
    Ni I &287  & 522 &  \cite{5298TP}, \cite{UlfLitzen_1993} \\
    Cu I & 360 &  37&  \cite{7574TP}, \cite{10.1063/1.555855} \\
    Zn I & 374 & 16 & \cite{7574TP}, \cite{10.1063/1.555971} \\
    Ga I & 257 &  23&  \cite{7574TP}, \cite{10.1063/1.2207144} \\
    Ge I & 618 & 26 &  \cite{4515TP}, \cite{10.1063/1.555929} \\
    As I & 84 & 14 & \cite{2177TP}, \cite{1987JPCRD..16S....K}  \\
    Br I & 264 & 54 & \cite{7285TP}, \cite{Humphreys:71} \\
    Kr I &527  & 184 &  \cite{7227TP}, \cite{10.1063/1.2227036} \\
    Rb I & 239 & 40 &  \cite{7857TP} \\
    Sr I & 377 & 86 & \cite{8792TP} \\
    Y I & 176 & 189 & \cite{9421TP}, \cite{10.1139/cjp-2016-0512} \\
    Mo I & 281 & 498 & \cite{5293TP}, \cite{10.1063/1.555818} \\
    Tc I & 274 &13  & \cite{7852TP}, \cite{PALMERI1999603} \\
    Ru I &  228& 11 & \cite{7852TP}, \cite{Callender:88} \\
    Rh I & 75 &111  & \cite{3464TP}, \cite{Callender:88} \\
    Pd I & 143 & 8 & \cite{7852TP}, \cite{RolfEngleman_1998} \\
    Ag I & 104 & 7 &  \cite{4515TP}, \cite{PhysRevA.74.062509} \\
    Cd I & 126 & 18 &  \cite{4515TP}, \cite{EVidolova-Angelova_1996} \\
    In I & 111 & 27 & \cite{7227TP}, \cite{HansKarlsson_2001} \\
    Sn I & 218 & 55 &  \cite{4515TP}, \cite{Brown:77} \\
    Sb I & 145 & 10 & \cite{7852TP}, \cite{Hassini:88} \\
    Te I &  119& 6 & \cite{7227TP}, \cite{YMakdisi_1982} \\
    I I &  220& 5 & \cite{7852TP}, \cite{10.1063/1.461601} \\
    Xe I & 443 & 187 & \cite{7227TP}, \cite{10.1063/1.1649348} \\
    Cs I & 178 & 42 & \cite{8682TP} \\
    Ba I & 355 & 109 &  \cite{4515TP},\cite{10.1063/1.1643404} \\
    La I & 206 &279  & \cite{9460TP}, \cite{Martin2023} \\
    Ce I & 354 & 54 &  \cite{8748TP}, \cite{Martin2023} \\
    Nd I & 229 & 4 & \cite{7852TP}, \cite{Martin2023} \\
    Sm I &220  & 7 &  \cite{7852TP}, \cite{Martin2023}  \\
    Eu I & 316 & 142 &  \cite{7227TP}, \cite{Martin2023} \\
    Gd I & 465 & 16 &  \cite{7852TP}, \cite{Martin2023} \\
    Dy I & 302 & 40 & \cite{4515TP}, \cite{7852TP} \\
    Ho I &  201& 12 & \cite{7852TP}, \cite{Martin2023} \\
    Er I & 313 & 10 &  \cite{7852TP}, \cite{Martin2023} \\
    Tm I & 477 & 349 & \cite{6708TP}, \cite{Martin2023} \\
    Yb I & 182 &5  &   \cite{7852TP}, \cite{Martin2023} \\
    Lu I & 198 & 44 &  \cite{7218TP}, \cite{JVerges_1978} \\
    Hf I & 278 & 187 &  \cite{5346TP}, \cite{Lawler_2022} \\
    Ta I & 122 & 135 &  \cite{7871TP}, \cite{Martin2023} \\
    W I & 144 & 121 &  \cite{7858TP} \\
    Ir I & 53 & 37 & \cite{7959TP}, \cite{Martin2023} \\
    Pt I & 184 & 166 &  \cite{3313TP}, \cite{refId0} \\
    Au I & 61 &  20&  \cite{7963TP}, \cite{7852TP} \\
    Hg I & 294 & 53 & \cite{3464TP}, \cite{10.1063/1.2204960} \\
    Tl I & 66 & 9 & \cite{4515TP}, \cite{Martin2023} \\
    Pb I & 134 & 28 &  \cite{7227TP}, \cite{Wood:68} \\
    Bi I &  71 & 39 &  \cite{8655TP}, \cite{Wahlgren_2001}\\
    Fr I & 122 & 149 & \cite{3464TP}, \cite{10.1063/1.2719251} \\
    Ra I & 81 & 19 & \cite{8693TP}, \cite{10.1063/1.4940416} \\
    Ac I & 52 & 91 & \cite{10463TP} \\
\end{longtable}

\begin{longtable}{crrl}
    \caption{\label{tb15} Summary of singly charged atoms (including isotopes)  with \texttt{.trans} and \texttt{.states} files from \textsc{NIST} including $N_{\text{state}}$, $N_{\text{trans}}$ and source references.} \\
    \toprule
    Element & $N_{\text{state}}$ &  $N_{\text{trans}}$ & References \\
    \midrule
    \endfirsthead

    \multicolumn{4}{c}%
    {{\bfseries \tablename\ \thetable{} -- continued from previous page}} \\
    \toprule
    Element & $N_{\text{state}}$ &  $N_{\text{trans}}$ & References \\
    \midrule
    \endhead

    \midrule \multicolumn{4}{r}{{Continued on next page}} \\
    \endfoot

    \bottomrule
    \endlastfoot
    $^{3}$He II & 148 &  140& \cite{8610TP}, \cite{10.1063/1.4927487} \\
    $^{4}$He II & 148 & 140 &  \cite{10.1063/1.1796671}, \cite{10.1063/1.4927487} \\
    Li II & 178 & 564 & \cite{8610TP}, \cite{10.1139/p88-100} \\
    Be II & 249 & 149 & \cite{8702TP}, \cite{AEKramida_2005} \\
    B II &156  & 435 &  \cite{8702TP}, \cite{ANRyabtsev_2005} \\
    C II & 414 &  1433&  \cite{10528TP} \\
    N II & 178 &786  &  \cite{6574TP}, \cite{SunHuLiLiuMeiGou+2021+1+12} \\
    O II & 275 & 876 &  \cite{6492TP}, \cite{10.1139/cjp-2016-0512} \\
    F II & 290 & 67 &  \cite{1115TP}, \cite{1987JPCRD..16S....K} \\
    Ne II & 380 & 233 &  \cite{1115TP}, \cite{892816} \\
    Na II &  162& 176 &  \cite{8098TP}, \cite{10.1063/1.2943652} \\
    Mg II & 137 & 482 &  \cite{10.1063/1.555617}, \cite{8098TP} \\
    Al II & 217 & 986 &  \cite{8199TP}, \cite{10.1063/1.555608} \\
    Si II & 148 & 474 & \cite{8363TP}, \cite{10.1063/1.555685} \\
    P II & 160 & 73 & \cite{1619TP}, \cite{10.1063/1.555736} \\
    S II & 232 & 753 &  \cite{8595TP}, \cite{10.1063/1.555862} \\
    Cl II & 274 & 221 & \cite{1619TP}, \cite{Radziemski:74} \\
    Ar II & 418 & 307 &  \cite{1619TP}, \cite{10.1063/1.3337661} \\
    K II & 96 & 252 &  \cite{8198TP} \\
    Ca II & 71 & 99 & \cite{1619TP}, \cite{osti_6008300}  \\
    Sc II &  168& 139 & \cite{5297TP}, \cite{osti_6008300} \\
    Ti II &  252& 470 & \cite{5297TP}, \cite{10.1063/1.3656882} \\
    V II & 407 &1884  & \cite{9770TP} \\
    Cr II & 913 & 92 &  \cite{7574TP}, \cite{osti_6008300} \\
    Mn II & 514 & 840 &  \cite{9140TP} \\
    Fe II & 1027 & 7293 &  \cite{7958TP}, \cite{Nave_2013} \\
    Co II & 478 & 2746 &  \cite{6999TP}, \cite{osti_6008300} \\
    Ni II & 713 & 208 &  \cite{5298TP}, \cite{osti_6008300} \\
    Cu II & 467 & 553 &  \cite{9697TP} \\
    Zn II & 93 & 22 &  \cite{7574TP}, \cite{10.1063/1.555971} \\
    Ga II & 95 & 10 &  \cite{10.1063/1.2207144} \\
    Ge II & 127 & 20 &  \cite{4515TP}, \cite{10.1063/1.555929}\\
    Kr II & 162 & 20 &  \cite{3464TP}, \cite{10.1063/1.2227036} \\
    Rb II & 165 & 49 &  \cite{7857TP} \\
    Sr II & 70 & 33 &  \cite{8986TP}  \\
    Y II & 235 & 66 &  \cite{4373TP}, \cite{AENilsson_1991}\\
    Tc II & 33 & 6 &  \cite{7852TP}, \cite{10.1139/cjp-2016-0512} \\
    Ru II & 225 & 8 &  \cite{7852TP}, \cite{HKarlsson_2002} \\
    Pd II &  185& 10 &  \cite{7852TP}, \cite{Lundberg_2001} \\
    Ag II & 99 & 236 &  \cite{9142TP} \\
    Cd II & 95 & 87 &  \cite{3315TP}, \cite{10.1139/cjp-2016-0512} \\
    In II & 194 & 528 & \cite{9131TP} \\
    Sn II &  76& 134 &  \cite{9352TP} \\
    Sb II & 109 & 2 &  \cite{7852TP}, \cite{10.1139/cjp-2016-0512}  \\
    Xe II & 161 & 22 & \cite{3464TP}, \cite{10.1063/1.1649348} \\
    Cs II & 315 & 2 & \cite{8682TP} \\
    Ba II & 161 & 83 &  \cite{4515TP}, \cite{10.1063/1.1643404} \\
    La II & 109 & 84 &  \cite{7455TP}, \cite{GUZELCIMEN2018188} \\
    Ce II & 480 &  283&  \cite{8629TP}, \cite{PhysRevA.91.022504} \\
    Pr II & 111 & 101 &  \cite{8137TP}, \cite{Radziute_2020} \\
    Nd II & 325 & 99 &  \cite{7685TP}, \cite{Radziute_2020} \\
    Sm II & 127 & 7 & \cite{7852TP}, \cite{Radziute_2020} \\
    Eu II & 57 & 13 &  \cite{7852TP}, \cite{Radziute_2020}  \\
    Tb II & 62 & 8 &  \cite{7852TP} \\
    Dy II & 300 & 17 & \cite{7852TP} \\
    Ho II & 31 & 4 &  \cite{7852TP}, \cite{Radziute_2021} \\
    Er II & 111 & 7 &  \cite{7852TP}, \cite{Radziute_2021}  \\
    Tm II &351  & 13 & \cite{7852TP}, \cite{Radziute_2021}  \\
    Yb II & 284 & 10 & \cite{7852TP}, \cite{Radziute_2021}  \\
    Lu II & 37 & 9 & \cite{7852TP}, \cite{PhysRevA.100.062505} \\
    Hf II & 106 & 2 &  \cite{7852TP}, \cite{PhysRevA.106.032807} \\
    W II & 58 &71  & \cite{7858TP} \\
    Ir II & 53 & 126 &  \cite{7959TP}, \cite{VANKLEEF1978251} \\
    Pt II & 277 & 183 &  \cite{8261TP}, \cite{Jean-FrancoisWyart_1995} \\
    Hg II & 114 & 446 &  \cite{7852TP}, \cite{Sansonetti_2001} \\
    Tl II & 76 & 3 &  \cite{7852TP} \\
    Pb II & 90 & 3 &  \cite{7852TP} \\
    Bi II &77  & 4 &  \cite{7852TP}, \cite{refId1} \\
    Ra II &  36& 9 &  \cite{8601TP}, \cite{10.1063/1.4940416}\\
    Ac II & 83 &  288 &  \cite{10463TP} \\

\end{longtable}

\newpage

\begin{table}
\centering
\caption{\label{tb6}$N_{\text{state}}$ and $N_{\text{trans}}$ for each Neutral in Kurucz Database
\citep{Kurucz2014,Kurucz_2009,Castelli_2010,Peterson_2014,Peterson_2022,KuruczNeed2002,Castelli_2015}}
\begin{tabularx}{\textwidth}{lrrX}
\toprule
Element & $N_{\text{state}}$ & $N_{\text{trans}}$ & References \\
\midrule
Li I & 123 & 1424 & \cite{CAVE,LN,1115TP,SEN} \\
Be I & 663 & 23574 & \cite{LCV,LV,1115TP,PST} \\
B I  & 1549 & 22012 & \cite{BRO,KP,MBO,1115TP} \\
C I  & 3500 & 250828 & \cite{BDEL,BRO,BRS,CM,GOLY,KP,LAMB,MS,MWRB,1115TP,NUSS,SMW,SW,6492TP} \\
N I  & 4479 & 356876 & \cite{BRS,DBG,KP,1115TP,NSW,6492TP} \\
O I  & 3932 & 109396 & \cite{BDEL,KP,1115TP,6492TP} \\
F I  & 1959 & 131375 & \cite{KP} \\
Ne I & 777  & 15361  & \cite{BM,KP,L75,L75,MC_2,MC_28,MC_29} \\
Na I & 3131 & 51288  & \cite{KP,LN,1619TP} \\
Mg I & 588  & 22612  & \cite{AZ,FF,FF2,FT,K75,KP,LV,LZ,1619TP,SCH,SL,WARN} \\
Al I & 1888 & 29081  & \cite{K75,KP,LC,1619TP,ROIG,SL} \\
Si I & 3177 & 171085 & \cite{GARZ,GS,K75,KP,MIL,SG} \\
P I  & 2387 & 202300 & \cite{KP,LAW,LKIP,MRB,1619TP} \\
S I  & 4171 & 13026  & \cite{FOS,KP,1619TP} \\
Cl I & 1727 & 255821 & \cite{KP} \\
Ar I & 801  & 49824  & \cite{AS,DC,KP,LI,1619TP,WA} \\
K I  & 2865 & 88919  & \cite{KP,MS_2,1619TP} \\
Ca I & 1127 & 46062  & \cite{BW,K88,1619TP,NEW,NEWS} \\
Sc I & 4307 & 724611 & \cite{FPAR,K88,5297TP,NBS_3} \\
Ti I & 13977& 5071484& \cite{K88,KLEM,LCJ,5297TP,NBS_3,SK,WB} \\
V I  & 29719& 7056409& \cite{K88,KING,5297TP} \\
Cr I & 37498& 2744900& \cite{BYAR,CSE,HS,HST,HT,K88,5297TP,NBS_4,WBCW} \\
Mn I & 37983& 1469883& \cite{BC,GW,K88,5297TP,NBS_4} \\
Fe I & 37504& 7498317& \cite{BH,BIPS,BK,FHSW,5298TP,GHP,HT,HUBR,HZ,K88,MRW,WBW2,WMML} \\
Co I & 23997& 3752335& \cite{FHSW,5298TP,K88,NBS_5,RMC} \\
Ni I & 10061& 732160 & \cite{5298TP,HEIS,K88,KING_2,LW,LWST,NBS_5,WL} \\
Cu I & 2178 & 32242  & \cite{BIEL,BIEM,CORL} \\
Zn I & 2212 & 129290 & \cite{AJG,AS_2,AZ_2,LMW,MOIS,BW} \\
Sr I & 1147 & 43222  & \cite{CB} \\
Y I  & 3768 & 586984 & \cite{CB,CP',5298TP} \\
Zr I & 14107& 5055470& \cite{BG,CB} \\
Nb I & 29578& 7170524& \cite{CB,DL} \\
Mo I & 38117& 3123467& \cite{CB,SD,5293TP,WH} \\
Tc I & 38320& 5090717& \\
Ru I & 36619& 7421138& \cite{BG_2,CB,SL_2} \\
Rh I & 24063& 3035030& \cite{CB,DL_2,KZ,SDL} \\
Pd I & 10229& 571621 & \cite{BG_3,CB} \\
Ba I & 1147 & 66443  & \cite{CB,NBS_6} \\
\bottomrule
\end{tabularx}
\end{table}

\begin{table}
\centering
\caption{\label{tb7}
Count of States and Transition Lines for each Singly Charged Ion in Kurucz Database}
\begin{tabularx}{\textwidth}{lrrX}
\toprule
Element & \texttt{N\_states} & \texttt{N\_trans} & References \\
\midrule
Be II & 123 & 1418 & \cite{LAU,1115TP} \\
B II  & 763 & 34046 & \cite{BRO,CJ,HIB,LCV,MBO} \\
C II  & 1569 & 28924 & \cite{BRO,CHY,DPD2,KP,LD,MC,1115TP,PDC,SIN} \\
N II  & 3144 & 169232 & \cite{DBG,KP,1115TP,NSW,TA} \\
O II  & 3737 & 296883 & \cite{KP,1115TP} \\
F II  & 4240 & 206549 & \cite{KP,1115TP,PILK} \\
Ne II & 1959 & 178364 & \cite{KP,LG,1115TP} \\
Na II & 777 & 52696 & \cite{KP} \\
Mg II & 324 & 6018 & \cite{BWL,KP} \\
Al II & 939 & 31420 & \cite{BW,KP,1619TP,WEI} \\
Si II & 1764 & 36511 & \cite{AJPP,BBCB,CS,KP,1619TP,SG} \\
P II  & 3254 & 217038 & \cite{HI,KP,MRB,1619TP} \\
S II  & 2387 & 275137 & \cite{BSMB,BW_2,KP,MWRB,1619TP} \\
Cl II & 3110 & 258102 & \cite{KP,1619TP} \\
Ar II & 2025 & 230953 & \cite{GO,KP,L_2,LG_2,1619TP,T} \\
K II  & 801 & 55355 & \cite{BBB,KP} \\
Ca II & 2884 & 115061 & \cite{BWL,K88,1619TP} \\
Sc II & 1153 & 116491 & \cite{K88,5297TP,NBS_3} \\
Ti II & 4414 & 897984 & \cite{CWAR,K88,5297TP,NBS_3,RVC,WB} \\
V II  & 14162 & 4347545 & \cite{CWAR,K88,5297TP,NBS_4} \\
Cr II & 29657 & 8683460 & \cite{BYAR,CWAR,K88,5297TP,NBS_4,SHAC} \\
Mn II & 39503 & 5124343 & \cite{K88,5297TP} \\
Fe II & 39418 & 7823961 & \cite{5298TP,K88} \\
Co II & 38019 & 7951292 & \cite{5298TP,K88} \\
Ni II & 21695 & 55833 & \cite{5298TP,GMW,HEIS,K88,MOIT} \\
Cu II & 10061 & 584247 & \cite{KP,KR} \\
Zn II & 2212 & 968 & \cite{AS_2,BS} \\
Sr II & 296 & 5993 & \cite{WA_2} \\
Y II  & 1990 & 162441 & \cite{CC,5298TP} \\
Zr II & 4021 & 1026414 & \cite{BG,CC} \\
Nb II & 14110 & 15047 & \cite{CB,HLB} \\
Mo II & 29596 & 13272 & \cite{CB,SD} \\
Tc II & 38246 & 119 &  \\
Ru II & 38321 & 14670073 & \cite{CB,MC_3} \\
Rh II & 36653 & 13817566 & \cite{CB} \\
Pd II & 24065 & 5191374 & \cite{CB} \\
Ba II & 119 & 1422 & \cite{NBS_6,WA_2} \\
\bottomrule
\end{tabularx}
\end{table}

\section{Database Structure}

The Django web framework is used for the database implementation of \textsc{ExoAtom}, written in the Python programming language \citep{django}. The \textsc{ExoAtom} data structure is an extension of the original ExoMol data structure \citep{jt548,jt939}, see  \citet{jt898}. This data structure is designed to offer a full description of the metadata for each file; it can be routinely used for both downloading and updating data through an application program interface (API), see \citet{jt939}.  

While the majority of the molecular data in ExoMol were generated in-house, atomic data are not natively produced by the ExoMol team. Therefore, the contents of \textsc{ExoAtom} have been taken from  existing atomic spectroscopic databases:  \textsc{NIST} and Kurucz. Then, to ensure consistency with ExoMol formatting, a systematic approach is taken to extract, process, and convert data from these sources. After having been scraped from the original databases, the atomic data for each element are then converted to the three  primary, ExoMol-formatted files: the \texttt{.states}  file, the transitions \texttt{.trans} file, and the partition function \texttt{.pf}  file. 

The following section provides a detailed review of the ExoMol format and precisely which data were scraped from the \textsc{NIST} and Kurucz databases to supply the atomic data for \textsc{ExoAtom}.

As specified in  Table \ref{tb16}, the major difference between the two datasets is that \textsc{NIST} aims for accuracy while Kurucz aims for
completeness. Which of these properties is more important will depend on individual use cases so, in contrast to the molecular line lists
provided by ExoMol, we do not recommend a dataset for each atom or ion. Instead this choice needs to be made by the user based on their
particular data requirements.  

\begin{table}
\caption{\label{tb16}Data Extracted from \textsc{NIST} \citep{ralchenko2020development,NISTweb,NIST_ASD} and Kurucz \citep{Kuruczweb, kurucz2011including} Databases.}
\centering
\renewcommand{\arraystretch}{1.2} % Adjust row height
\setlength{\tabcolsep}{4pt} % Adjust column spacing
\begin{tabular}{l l l l l}
\hline
\textbf{Database} & \textbf{Levels} & \textbf{Lines} & \textbf{Scope} & \textbf{Characteristic} \\
\hline
\textsc{NIST}   & 94 neutral from H to Pa & 94 neutral from H to Pa & General applications & Accurate\\
       & 91 singly charged from He to Pa & 87 singly charged (He to Pa except Po, At, Rn, Fr) & & \\
Kurucz &  38 neutral \& 37 singly charged & 38 neutral \& 37 singly charged  & Hot stars & Complete\\
\hline
\end{tabular}
\end{table}

\subsection{ExoMol Format}

The ExoMol database consists of 15 types of files, as summarized in Table \ref{tb1}.
\begin{table}
\centering
\caption{\label{tb1}Specification of the ExoMol file types. (Contents in brackets are optional.)}
\begin{tabular}{lp{4cm}p{10cm}}
\toprule
File Extension & File DSname & Contents \\
\midrule
\texttt{.all}        & Master               & Single file defining contents of the ExoMol database. \\
\texttt{.def}        & Definition           & Defines contents of other files for each isotope. \\
\texttt{.states}     & States               & Energy levels, quantum numbers, uncertainties, lifetimes, (Land\'{e} $g$-factors). \\
\texttt{.trans}      & Transitions          & Einstein $A$ coefficients, (wavenumber). \\
\texttt{.broad}      & Broadening           & Parameters for pressure-dependent line profiles. \\
\texttt{.cross}      & Cross sections       & Temperature or temperature and pressure-dependent cross sections. \\
\texttt{.kcoef}      & $k$-coefficients     & Temperature and pressure-dependent $k$-coefficients. \\
\texttt{.pf}         & Partition function   & Temperature-dependent partition function. \\
\texttt{.cf}         & Cooling function     & Temperature-dependent cooling function. \\
\texttt{.cp}         & Specific heat        & Temperature-dependent specific heat. \\
\texttt{.super}      & Super-lines          & Temperature-dependent super-lines (histograms) on a wavenumber grid. \\
\texttt{.nm}         & Vacuum-Ultraviolet (VUV) cross sections   & Temperature and pressure-dependent VUV cross-sections (wavelength, nm). \\
\texttt{.fits}, \texttt{.h5}, \texttt{.kta} & Opacities        & Temperature and pressure-dependent opacities for radiative-transfer applications. \\
\texttt{.overview}   & Overview             & Overview of datasets available. \\
\texttt{.readme}     & Readme               & Specifies data formats. \\
\texttt{.model}      & Model                & Specification of the spectroscopic model. \\
\bottomrule
\end{tabular}
\end{table}

The \textsc{ExoAtom} database selects the \texttt{.all}, \texttt{.def}, \texttt{.states}, \texttt{.trans}, and \texttt{.pf} files from ExoMol as fundamental components for structuring atomic data (see Table \ref{tb2}). The \texttt{.states} file contains energy levels along with their quantum numbers, uncertainties, lifetimes, term, etc. Optionally the \texttt{.states} file can provide Land\'e $g$-factors which can be used to give the splitting of levels in weak magnetic fields.  For each transition, the \texttt{.trans} file provides identifiers to the two states involved, the associated Einstein $A$ coefficient and corresponding wavenumbers. The \texttt{.pf} file provides partition functions over a range of temperatures.

\begin{table}
\caption{\label{tb2}Specification of the \textsc{ExoAtom} database file types.}
\centering
\begin{tabular}{lcll}
\hline
\textbf{File extension} & \textbf{$N_{\rm files}$} & \textbf{File DSname} & \textbf{Contents} \\
\hline
.all   & 1 & Master               & Single file defining contents of the \textsc{ExoAtom} database \\
.adef.json  & $N_{\rm iso}$ & Definition           & Defines contents of other files for each isotope \\
.states& $N_{\rm iso}$ & States               & Energy levels, quantum numbers, uncertainties,\\
       &   &                      & lifetimes, (Land\'e g-factors) \\
.trans & $N_{\rm iso}$ & Transitions          & Einstein A coefficients (wavenumber) \\
.pf    & $N_{\rm iso}$ & Partition Functions  & Parameters for pressure-dependent line profiles \\
\hline
\multicolumn{4}{l}{$N_{\rm iso}$: total number of isotopes considered for each atom or ion (per data set).}\\
\end{tabular}
\end{table}

\subsection{\textsc{NIST} Databases}

The \textsc{NIST} Atomic Spectra Database was accessed via the \href{https://www.nist.gov/pml/atomic-spectra-database}{NIST Atomic Spectra Database} website \citep{ralchenko2020development,NISTweb,NIST_ASD}, and the data were scraped by \citet{Rujia} in June 2024. This database is essentially a compilation
of available, high accuracy experimental measurements which,  up to and including 2024, was updated annually.

 The \textsc{NIST} database consists of four primary sections: Lines, Levels, Ground States \& Ionization Energies, and Laser-Induced Breakdown Spectroscopy (LIBS). Data is extracted from the \textsc{NIST} Atomic Spectra Database Lines section and the Levels section for both neutral and singly charged atoms. The \texttt{.states} files are constructed from the energy levels data, while the \texttt{.trans} files incorporate both levels and spectral lines data. The \texttt{.pf} files are derived from the levels data.
 The \textsc{NIST} database includes lifetime data for some atomic states but lacks Land\'{e} $g$-factors.
  Land\'{e} $g$-factors are optional in the ExoMol data structure and are not provided for the \textsc{NIST} datasets. These quantities are important because they enable predictions of Zeeman splitting of rovibronic lines in the presence of magnetic fields, providing a remote-sensing probe of astrophysical environments.

 Uncertainties (Unc) of the energy levels  of the \textsc{NIST} data were determined as follows.  Generally, the uncertainty would be taken directly from \textsc{NIST}. If the \textsc{NIST} uncertainty value `Unc' was absent (`\_' or blank), it was estimated  based on the number of decimal places ($N$) in the Level dataset, using the formula \(\text{Unc} = 2 \times 10^{-N}\)~cm$^{-1}$. For example, if Level \((\mathrm{cm}^{-1}) = 12345.24\), then the corresponding Unc is 0.02 . If Level \((\mathrm{cm}^{-1})\) contains a decimal point but no decimal places (e.g., `74728.'),  Unc is set to 2.  
 
 %\pink{QH:"Some of the energy levels in ASD do have uncertainties (e.g., Fe I and II), and all new data compilations have energy uncertainties without exception. However, the older data may not specific uncertainties as you correctly noticed. Generally, we claim that it is between 2.5 and 25 units of the last digit."(different from original data)}

%\red{sy: again, too optimistic IMO.}
%\red{sy: From my experience, this is too optimistic estimate of uncertainty. The last decimal place in 12345.24 usually meas that the associated but absent uncertainty is within 0.09. Unless we were instructed to do so by NIST, I would go for 0.1-0.09 cm-1}
%NIST taken uncertanty direct;y

Due to the absence of valid Einstein $A$ coefficients for certain radioactive elements, the \texttt{.trans} files scraped from \textsc{NIST} are absent for elements such as Polonium (Po), Astatine (At), Radon (Rn), and Francium (Fr). These elements exhibit short half-lives, making experimental measurements particularly challenging and limit the likelihood of their astronomical detection. However, \texttt{.states} and \texttt{.pf} files are available for these atoms. 

%In some cases the lines  containing valid \(A_{ki}\) (\(\mathrm{s}^{-1}\))  lacks valid $E_i$, $E_k$, $\text{conf}_k$, or $\text{conf}_i$ in the corresponding rows thus making their correlation to the states impossible. In these cases, the Einstein $A$ coefficients were excluded. %\red{sy: I have moved and rephrased the previous sentence from the Transition section. I think it belongs here. Please check if I interpretted it correctly. }

%absent
In some cases, the lines containing valid \(A_{ki}\) (\(\mathrm{s}^{-1}\)) lack valid 
\(E_i\), \(E_k\), configuration labels (\(\text{conf}_i\), \(\text{conf}_k\)), or terms 
in the corresponding rows, thus making their correlation to the states impossible. 
In these cases, the Einstein \(A\) coefficients were excluded.

%\red{sy: what about the state degeneracy? Is is provided in NIST? As 2J+1? If so, I think this place would be appropriate to introduce the alternative definition of $g_J$.} 

%\red{sy: I have mode the discussion of the extraction of NIST partition functions from below to this section. I think it belongs here}

The \textsc{NIST} database provides partition functions within individual HTML files, but each file only allows retrieval for a single temperature. Consequently, obtaining partition functions across a temperature range from 1 K to 6000 K requires extracting data from 6000 separate documents. 
We have therefore computed these partition function by  direct summation, see discussion below.

\subsection{Kurucz Databases}%(?)

The Kurucz Atomic Line Lists were accessed via the \href{http://kurucz.harvard.edu/atoms.html}{Kurucz Atomic Database} website \citep{Kuruczweb, kurucz2011including} and the data were scraped by \citet{tianyang} in June 2024. The Kurucz database appears to have been last updated in 2017.

In the Kurucz database each neutral atom and singly charged ion are designated by a four-digit identifier (\texttt{xxyy}), where \texttt{xx} corresponds to the atomic number and \texttt{yy} indicates whether the species is neutral (\texttt{yy = 00}) or singly charged (\texttt{yy = 01}). The  Kurucz line lists are calculated
although some energy data has been replaced by values taken (at an unspecified date) from \textsc{NIST}; these updates are marked in 
the database.The Kurucz Database does not provide the $A$ coefficients directly 
but instead includes $\log_{10} A$ values in the \texttt{.agafgf} files.
  For certain atoms (S I, Ni II, Nb II, Mo II), the  Kurucz atomic \texttt{lines} files are not available within the specified path. The Kurucz database provides Land\'{e} $g$-factors (stored in files gfxxyy.gam) and  lifetimes (stored in lifexxyy.dat or gfxxyy.life files). 
However, these lifetime values are not always complete or consistently available; missing values were therefore  recomputed using the program  \textsc{PyExoCross} \citep{jt914}

 % but does not include lifetime data; the  lifetimes were therefore  computed using program \textsc{PyExoCross} \citep{jt914} and integrated with the available Land\'e $g$-factors (see below and Table \ref{tb16}). 

\begin{table}
\centering
\caption{\label{tbbb}Source file for the value in \texttt{.trans} file of Kurucz Database.}
\begin{tabular}{ll}
\toprule
Source File Name & Values Contained \\
\midrule
gfxxyy.lines &   $i$, $f$   \\
gfxxyy.agafgf  & $A$, $\tilde{\nu}_{if}$  \\
\bottomrule
\end{tabular}
\end{table}

Table \ref{tbbb} shows which files from the Kurucz Database are the source to form the ExoMol atomic \texttt{.trans} files, while  Table \ref{tb3} shows which files from the Kurucz Database are the source to form the ExoMol atomic \texttt{.states} files.  Although the Kurucz database contains several different files for transition lines, the \textsc{NIST} database provides more accurate data. To ensure a more complete dataset, both predicted and measured lines from the Kurucz database are included. The \texttt{gfxxyy.lines} files are selected as they contain the majority of transition lines. However, the \textit{A} coefficient and wavenumbers are not provided within the \texttt{.lines} files, for which we scraped the \texttt{.agafgf} files, which contain both values. The \texttt{lines} files do not have indices directly. Therefore, Energy, \(J\), Configuration, and Term are matched to the \texttt{.states} file to retrieve the index.

 To obtain the states data, there are only two files that hold the necessary values. One is \texttt{AELxxyy.DAT} and the other is \texttt{gfxxyy.gam}. However, the \texttt{AELxxyy.DAT} is now obsolete and contains less data than \texttt{.gam} files, which have all energy levels with $J$, $QN$ and $g$. Therefore, \texttt{gfxxyy.gam} was chosen to be scraped for the states. Lifetimes are stored in files   \texttt{lifexxyy.dat} and \texttt{gfxxyy.life}, but the contents are the same. Uncertainties and state degeneracies, $g_{J}$, are not provided by the Kurucz database. Uncertainties for the Kurucz energy levels were arbitrarily set to 0.1 cm$^{-1}$ and $g_{J}$ calculated as $2J+1$. %\red{this is the first time the state degeneracy is introduced as 2J+1, i.e. with no nuclear spin factor, which is different from the molecular ExoMol. }. 
% \red{sy: How many decimal places are used for Kurucz's energy and transition values? I think this would be an obvious question to ask by a reviewer and to wonder by a reader.}
 %\red{sy: I am confused, the previous paragraph says Kurucz "does not include lifetime data" which contradicts the latter sentence}

\begin{table}
\centering
\caption{\label{tb3}Source file for the value in \texttt{.states} file of Kurucz Database}
\begin{tabular}{ll}
\toprule
Source File Name & Contained Values \\
\midrule
gfxxyy.gam & $\hat{E}$, $J$, $g$, Configuration, Term     \\
lifexxyy.dat/gfxxyy.life  & $\tau$  \\
Computed & $i$, $g_{J}$, Abbr\\
\bottomrule
\end{tabular}
\end{table}

For some species (Na I, K I, Ca II, Zn I, Y II), the Kurucz database provides three versions of the same file with different suffixes (w, y, z), but the data we scraped from all three files are identical. The Kurucz database also contains information on radiative, Stark and Van der Waals damping for each species. These data are not included in the present version of \textsc{ExoAtom}. %\red{sy: were these  duplicates identical, to the last digit? If not, which values were prioritized? This is a technical question though.}

%\pink{In the generated states files, only the necessary columns from the Kurucz GAM files are retained. For these retained columns, entries from different suffix files (w, y, z) are identical down to the last digit. Although the original GAM files include additional columns that may differ slightly, those fields are not utilized by the states-generation pipeline and were therefore ignored during deduplication+.}

%\red{sy: This paragraph was moved from the ExoMol Partition function section. I think it belongs here}

In contrast to \textsc{NIST}, the Kurucz database provides partition functions in a single file, \texttt{partfnxxyy.dat}, which contains multiple partition function values under different potential lowering conditions, expressed in \(\mathrm{cm}^{-1} / Z_{\mathrm{eff}}^2\). These conditions are represented by column headers with values of -500, -1000, -2000, -4000, -8000, -16000, and -32000. However, there is no clear documentation of this or criterion for selecting the appropriate columns. Partition functions from the first column were chosen and were converted to the ExoMol format. 

\subsection{File Naming Convention}
The file naming convention in the \textsc{ExoAtom} database ensures unique, descriptive, and machine-readable file names for each atom. The master file, named \texttt{ExoAtom.all.json}, defines the entire content of the \textsc{ExoAtom} database. All other files follow a standardized naming convention that applies to files with the \texttt{.adef.json}, \texttt{.states}, \texttt{.trans}, and \texttt{.pf} extensions. Each file name uses the format: \texttt{<atom\_slug>\_\_<database\_name>}.

The atom slug is defined as follows. For neutral atoms, the element symbol is used; for singly charged atoms, the element symbol is appended with ``\texttt{\_p}''. Note that  the \textsc{NIST} database has isotopically-resolved data for the neutral hydrogen atom (H, D, T) and singly-charged helium ion ($^3$He, $^4$He) ion. For these two elements, the mass number is placed before the element symbol, for example ``\texttt{2H}'' and ``\texttt{4He\_p}''. There are no isotopically-resolved data in the Kurucz Database. 

For example, the file name for a neutral iron atom from the \textsc{NIST} database is \texttt{Fe\_\_NIST}, while the file name for a singly charged iron atom is \texttt{Fe\_p\_\_NIST}. Hydrogen has three isotopes in \textsc{NIST}: protium, deuterium, and tritium. Each isotope is named with its mass number as a superscript preceding the H. The corresponding file names are \texttt{1H\_\_NIST}, \texttt{2H\_\_NIST}, and \texttt{3H\_\_NIST}, respectively.

\subsection{Data Format}

\subsubsection{The Definition File}

The \textsc{ExoAtom} database can be accessed at \url{https://exomol.com/exoatom}. 
 The core information about each atomic species in the \textsc{ExoAtom} database is contained within its JSON definition file. JSON (JavaScript Object Notation) is a lightweight, human-readable format for data exchange that is commonly used to organize and transmit its structured scientific data \citep{json}.
 
The atomic definition file is called \texttt{adef.json} and adheres to the \textsc{ExoAtom} format \texttt{<AtomSlug>\_\_<DatasetName>.adef.json}.  This file specifies the available atomic data for a given species and describes potential applications, see Table \ref{jsontable}. Appendix \ref{appendix:json_example} presents the definition file \texttt{adef.json} for  \textsuperscript{1}H  from the \textsc{NIST} dataset as a typical example. These standardized fields allow consistent identification and reference across various isotopes. The purpose of this definition file is outlined as follows.

\begin{table}
    \centering
    \caption{\label{jsontable}Specification of the \textsc{ExoAtom} Definition File. }%\red{Why no mention of lifetimes in states section?}}
    \begin{tabularx}{\textwidth}{l X}
        \toprule
        \textbf{Field} & \textbf{Description} \\
        \midrule
        \multicolumn{2}{l}{\textbf{Species Information}} \\
        atom & Atomic symbol \\
        ordinary\_formula & Ordinary chemical formula \\
        spectroscopic\_notation & Spectroscopic notation \\
        charge & Charge of the species \\
        name & Name of the species \\
        mass\_in\_Da & Molecular mass in Daltons (Da) \\
        \\
        \multicolumn{2}{l}{\textbf{Isotope Information}} \\
        iso\_formula & Isotopic chemical formula \\
        iso\_name & Isotope name \\
        mass & Isotopic mass in Daltons (Da) \\
        spin & Nuclear spin value \\
        atomic mass number & Mass number of the isotope \\
        \\
        \multicolumn{2}{l}{\textbf{Dataset Information}} \\
        name & Dataset name \\
        version & Dataset version (YYYYMMDD format) \\
        doi & Digital Object Identifier for the dataset \\
        max\_temperature & Maximum temperature in dataset (K) \\
        n\_L\_default & Default number of Lorentzian pressure-broadening parameters value\\
        num\_pressure\_broadeners & Number of pressure broadeners \\
        nxsec\_files & Number of cross-section files \\
        nkcoeff\_files & Number of k-coefficient files \\
        dipole\_available & Availability of dipole data \\
        cooling\_function\_available & Availability of cooling function data \\
        specific\_heat\_available & Availability of specific heat data \\
        Ionisation & Ionisation data (null if not available) \\
        \\
        \multicolumn{2}{l}{\textbf{States Information}} \\
        number\_of\_states & Total number of states \\
        max\_energy & Maximum energy in dataset (cm$^{-1}$) \\
        uncertainty\_available & Indicates if uncertainty is described \\
        lifetime\_available & Availability of lifetime data \\
        lande\_g\_available & Availability of Land\'{e} $g$-factor \\
        num\_quanta & Number of quantum numbers \\
        states\_file\_fields & List of fields in the states file, including: \\
        \quad ID & Unique integer identifier for the energy level \\
        \quad E & State energy in cm$^{-1}$ \\
        \quad gtot & State degeneracy \\
        \quad J & Total angular momentum quantum number (integer/half-integer) \\
        \quad Unc & Uncertainty in the state energy in cm$^{-1}$ \\
        \quad gfactor & Land\'{e} $g$-factor (optional) \\
        \quad qn:configuration & Configuration for the state \\
        \quad qn:$LS$Coupling & Term for the state \\
        \quad qn:parity & Parity for the state \\
        \\
        \multicolumn{2}{l}{\textbf{Transitions Information}} \\
        number\_of\_transitions & Total number of transitions \\
        number\_of\_transition\_files & Number of transition files \\
        max\_wavenumber & Maximum wavenumber (cm$^{-1}$) \\
        transitions\_file\_fields & List of fields in the transition file, including: \\
        \quad i & Upper state ID \\
        \quad f & Lower  state ID \\
        \quad A & Einstein A coefficient (s$^{-1}$) \\
        \quad Wavenumber & Transition wavenumber in cm$^{-1}$ \\
        \\
        \multicolumn{2}{l}{\textbf{Partition Function}} \\
        max\_partition\_function\_temperature & Maximum temperature for partition function (K) \\
        partition\_function\_step\_size & Step size for partition function (K) \\
        fields & List of fields for the partition function, including: \\
        \quad T & Temperature in Kelvin \\
        \quad Q(T) & Partition function (dimensionless) \\
        \bottomrule
    \end{tabularx}
\end{table}

\textbf{Standardized \textsc{ExoAtom} File Usage}
The JSON structure follows a standardized format to ensure the identification for various atomic species. Each dataset follows predefined fields that enable structured organization, facilitating efficient data retrieval and comparison.

Within this definition file, information is organized into clearly defined sections that detail specific atomic characteristics and spectroscopic properties. The general JSON structure includes sections for Species, and Dataset (Dataset information; States, Transitions, and Partition Functions files information).

%The Species section provides basic atomic identifiers such as atomic symbol, ordinary and spectroscopic notation, ion charge, and atomic mass. The Dataset Section includes the general information about the dataset, including the dataset's name, version, maximum temperature, etc.  This section includes information about the three files for atomic data: states, transitions and partition function. The States section includes detailed energy-level data, providing unique identifiers, energies (in cm$^{-1}$), total angular momentum quantum numbers ($J$), uncertainties (cm$^{-1}$), degeneracies, electronic configurations, term symbols, and parity. The Transitions section details radiative transitions, specifying upper and lower energy state identifiers, Einstein $A$ coefficients (s$^{-1}$), and transition wavenumbers (cm$^{-1}$). The Partition Functions section offers temperature-dependent partition functions.

\textbf{Enhanced Database Functionality}
When isotopic data is provided for \textsc{ExoAtom}, the JSON file includes an additional Isotope section. This section describes the isotopic formula, isotope name, atomic mass, nuclear spin, and atomic mass number, see Table \ref{jsontable}. Nuclear spin ($I$) describes the intrinsic angular momentum of atomic nuclei, determined by the pairing of protons and neutrons. Different isotopes of an element have different numbers of neutrons, which affects their nuclear spin values. The nuclear spin depends on the total number of protons and neutrons and follows specific rules.  Including more complete isotope-resolved data and hyperfine-resolved transitions would enhance the database’s precision.

\textbf{Version Control and Dataset Updates}
Each dataset is version-controlled, ensuring that users can track modifications and access the latest available data. The \texttt{version} field, formatted as \texttt{YYYYMMDD}, allows systematic updates without requiring manual intervention. This versioning mechanism supports seamless data integration and enhances the reliability of long-term scientific studies.

\subsubsection{The Master File}

A summary file, named \texttt{exoatom.all.json}, consolidates the contents of the entire database. This file, which is available at \url{www.exomol.com/exoatom/exoatom.all.json}, provides a computer-readable (JSON format) list of recommended datasets, including those for each atomic species in the \textsc{ExoAtom} structure.

The master file summarizes the database's content and provides easy access to the latest version number of each dataset, allowing users to track updates efficiently. It begins with general information about the database, including the total count of atomic species, isotopes, and datasets sourced from three different projects. Specifically, this database currently includes 151 atomic species  from 2 databases.

Each atomic species in the \textsc{ExoAtom} master file is recorded separately, even if they correspond to a neutral atom and its ionized form. For example, iron (\(\text{Fe}\)) and its singly ionized state (\(\text{Fe}^+\)) are both treated as distinct entries, see Table~\ref{tab:exoatom_master_general}. Each entry includes the chemical formula, number of isotopes considered, as well as dataset sources and version numbers. This structure ensures that users can clearly distinguish between neutral atoms and their ionized states while maintaining consistency across datasets.

For elements with isotopic variations, each isotope is listed, as illustrated by  hydrogen and its
three isotopes three isotopes (\(^1\)H, \(^2\)H, \(^3\)H)
given in Table~\ref{tab:exoatom_master_isotope}. The master file provides detailed information, including the isotopic formula, dataset source, and version number, see Appendix \ref{appendix:master}. 

Table~\ref{tab:exoatom_master_general} and Table~\ref{tab:exoatom_master_isotope} illustrate the structured representation of atomic species in the master file. Table~\ref{tab:exoatom_master_general} presents general atomic species, using Fe and Fe\(^+\) as examples, where both the neutral and ionized forms are recorded separately. Table~\ref{tab:exoatom_master_isotope} demonstrates an element with multiple isotopes, using hydrogen as an example, detailing its three isotopic variants (\(^1\)H, \(^2\)H, and \(^3\)H) along with their dataset sources and version information.

\begin{table}
    \centering
    \caption{Extract from the \textsc{ExoAtom} Master file showing  general atomic species (Fe and Fe$^+$).}
    \label{tab:exoatom_master_general}
    \begin{tabular}{ll}
        \hline
        \textbf{Field} & \textbf{Description} \\
        \hline
        exoatom.master & ID \\
        20240601 & Version number (format YYYYMMDD) \\
       % \hline
        \multicolumn{2}{l}{\textbf{General Atomic Species (Iron)}} \\
        Atom Name & Iron \\
        Chemical Formula & Fe \\
        Number of Isotopes & 1 \\
        Dataset Name & \textsc{NIST}, Kurucz \\
        Version Number & 20240601 \\
        \multicolumn{2}{l}{\textbf{Iron Ion (II)}} \\
        Atom Name & Iron Ion (II) \\
        Chemical Formula & Fe\_p \\
        Number of Isotopes & 1 \\
        Dataset Name & \textsc{NIST}, Kurucz \\
        Version Number & 20240601 \\
        \hline
    \end{tabular}
\end{table}

\begin{table}
  \centering
    \caption{Extract from the \textsc{ExoAtom} Master File showing an atom with isotopes specified (hydrogen).}
    \label{tab:exoatom_master_isotope}
  \begin{tabular}{@{}ll@{}}
    \toprule
    \textbf{Field}       & \textbf{Description}                        \\
    \midrule
    name                 & Hydrogen                                    \\
    formula              & H                                           \\
    num\_isotopes        & 3                                           \\
    \multicolumn{2}{@{}l@{}}{\textbf{isotopes}}          \\
    iso\_slug            & \texttt{1H}, \texttt{2H}, \texttt{3H}                                  \\
    iso\_formula         & (1H), (2H), (3H)                            \\
    dataset              & NIST                                        \\
    version              & 20240901                                    \\
    \bottomrule
  \end{tabular}
\end{table}

\subsubsection{States Files}

\texttt{.states} files consist of the index $i$, energy term value $\tilde{E}$ (cm$^{-1}$),  the $J$-dependent state degeneracy $g_J$, quantum number $J$, uncertainty $\text{Unc}$ (cm$^{-1}$), lifetime $\tau$, Land\'e $g$-factor $g$, and state label $QN$ \citep{jt939}. 

%\texttt{.states} files consist of the index $i$, energy term value $\tilde{E}$ (cm$^{-1}$),  the $J$-dependent state degeneracy $g_J$ \red{sy: It is not total, but 2J+1 is it?}, quantum number $J$, uncertainty $\text{Unc}$ (cm$^{-1}$), lifetime $\tau$, Land\'e $g$-factor $g$, and state label $QN$ \citep{jt939}. 

For the \textsc{ExoAtom} database we have decided to follow the \textsc{NIST} convention and omit the nuclear spin degeneracy from $g_J$. This is different from the molecular ExoMol database which follows the HITRAN \citep{jt981} convention, which includes the nuclear spin degeneracy. This so-called physicist's convention is not adopted by  \textsc{ExoAtom} since for nearly all atoms in the database the isotope is not specified (astrophysicist's convention \citep{jt799}) and hence the value of the nuclear spin is not specified either. This means that the degeneracy factor $g_i$ is simply given by:
\begin{equation}
    g_i = 2J_i + 1
\end{equation}
where $J_i$ is the total angular momentum quantum number of $i^{\rm th}$ level.

In atomic spectroscopy, different coupling schemes describe how angular momenta combine to determine the total angular momentum $\vec{J}$. In \textsc{ExoAtom}, according to \citet{NISTterm}, the $LS$ coupling (Russell-Saunders coupling), $jj$ coupling (individual total angular momentum coupling), and Racah symbols are selected as the term types for the \textsc{NIST} Database. For the Kurucz Database, while essentially based on the $LS$-coupling, much of the notation  deviates from the standardized conventions typically used in spectroscopic databases. However, the original notation has been retained in its current form for the time being, with a potential revision considered as part of a future upgrade. Therefore, in all \texttt{JSON} files of the Kurucz Database, the corresponding value for the term field is simply set to ``Kurucz''.

The $LS$ coupling is used mostly in light atoms where the spin-orbit coupling is weak. Firstly, the total orbital angular momentum $\vec{L}$ and total spin angular momentum $\vec{S}$ of all electrons couple, forming the total angular momentum 
$ \vec{J}  =  \vec{L}  + \vec{S}$. 
%$\lvert \vec{J} \rvert = \lvert \vec{L} \rvert + \lvert \vec{S} \rvert$. 
The resulting terms follow the spectroscopic notation ${}^{2S+1}L_J$, where $S$ represents the total spin multiplicity, $\vec{L}$ is the total orbital angular momentum (denoted by spectroscopic symbols such as $S, P, D, F,\dots$), and $\vec{J}$ is the total angular momentum. In contrast, \textit{$jj$} coupling is more appropriate for heavy atoms where spin-orbit interaction is strong. Here, each electron’s individual orbital angular momentum $\vec{l}$ and spin $\vec{s}$ couple first to form $\vec{j} = \vec{l} + \vec{s}$, and then these $\vec{j}$ values combine to determine the total angular momentum $\vec{J}$ of the system. The resulting terms are written in terms of $\vec{j}$-values rather than $\vec{L}$ and $\vec{S}$, making this coupling more relevant for relativistic calculations. Racah symbols are used to further analyze angular momentum coupling. These symbols, denoted as $\langle j_1\, j_2 \mid j_3\, j_4 \rangle$, describe recoupling coefficients that simplify angular momentum calculations, such as Clebsch-Gordan and Racah coefficients. %These methods are basic in atomic structure calculations, particularly for complex electron configurations.

Table \ref{tb8} defines the specification of  \textsc{ExoAtom} \texttt{.states} file both for data from \textsc{NIST} Database and Kurucz Database. It includes key spectroscopic parameters such as the state index, energy, degeneracy, total angular momentum $J$, uncertainty, Land\'{e} $g$-factor, lifetime, and relevant configuration or term labels. Table \ref{tb20} and \ref{tb9} illustrates a representative sample (Li I) \texttt{.states} file from \textsc{NIST} Database and Kurucz Database, respectively.
The value ``Abbr'', see Table \ref{tb3},  is used for the Kurucz data to distinguish whether a state is predicted or not: predicted/calculated energies are labeled "CA", while measure empirical energies taken from \textsc{NIST}, are labeled "NI"; see \citet{jt948} for a discussion of these tags.  Differences between the ``NI'' levels in Kurucz and those listed in the \textsc{NIST} \texttt{.states} file arise because the Kurucz dataset is a hybrid compilation that combines both observed and theoretical energies. 
Therefore, the retained \textsc{NIST} identifiers indicate the observational origin of the term assignment rather than ensuring identical numerical energy values.

\begin{table}
\centering
\caption{Specification of the \textsc{ExoAtom} \texttt{.states} file.}
\label{tb8}
\begin{tabular}{l l l l}
    \hline
    \textbf{Field} & \textbf{Fortran format} & \textbf{C format} & \textbf{Description} \\
    \hline
    ID & \texttt{I12} & \texttt{\%12d} & State ID \\ 
    $\tilde{E}$ & \texttt{F12.6/} & \texttt{\%12.6f/} & State energy in \(\mathrm{cm}^{-1}\) \\ 
    & \texttt{F12.5/F12.4} & \texttt{\%12.5f/\%12.4f} &  \\
    $g_{J}$  & \texttt{I6} & \texttt{\%6d} & State degeneracy \\ 
    $J$ & \texttt{I7/F7.1} & \texttt{\%7d/\%7.1f} & Total angular momentum quantum number, $J$ (integer/half-integer) \\ 
    \text{Unc} & \texttt{F12.6} & \texttt{\%12.6f} & Uncertainty in the state energy in \(\mathrm{cm}^{-1}\) \\ 
    $\tau$$^\dagger$ & \texttt{ES12.4} & \texttt{\%12.4e} & Radiative lifetime in s (optional, Kurucz only) \\
    gfactor$^\dagger$ & \texttt{F10.6} & \texttt{\%10.6f} & Land\'{e} $g$-factor (optional, Kurucz only) \\ 
    qn:configuration & \texttt{A12}& \texttt{\%12s} & Configuration for the state \\
    term &\texttt{A8} & \texttt{\%8s} & Term for the state\\
    qn:parity$^\dagger$ & \texttt{A1}& \texttt{\%1s} & Parity for the state (optional, NIST only) \\
    Abbr$^\dagger$ & \texttt{A2} & \texttt{\%2s} & Abbreviation indicating data source: \texttt{CA} (calculated, Kurucz) or \texttt{NI} (measured, NIST) \\
    \hline
\end{tabular}
\vspace{10pt} % Add some vertical space before the multicolumn note
\noindent\begin{minipage}{\textwidth}
\footnotesize
\textbf{Note:} The \texttt{.states} file typically contains 9--10 columns. Columns marked with $^\dagger$ are optional, depending on the data source. 
The \texttt{.states} file is generated from the levels data using the following conversion rules:
\begin{itemize}
\item{The ExoMol data standards use the given data formats for columns which should be separated by a single space.}
    \item ID: Integer, starts from 1 and ends with the number of valid rows in levels data.
    \item $\tilde{E}$: Different formats apply: 
    $\tilde{E} \leq 100000$: \texttt{F12.6 or \%12.6f}, $100000 \leq \tilde{E} < 1000000$: \texttt{F12.5 or \%12.5f}, $\tilde{E} \geq 1000000$: \texttt{F12.4 or \%12.4f}.
    \item $g_{J}$: Obtained directly from $g$ in levels data. When the atom has isotopes, it corresponds to $g_{\text{tot}}$; when the atom does not have isotopes, it corresponds to $g_{J}$.
    \item $J$: Obtained directly from $J$ in levels data.
    \item Unc: Obtained directly from Uncertainty (\(\mathrm{cm}^{-1}\)) in levels data.
    \item $g$: Obtained from Land\'{e} $g$-factor column in levels data. If this column is absent, it will not appear in the \texttt{.states} file.
    \item qn:configuration: Directly from the configuration column in levels data. The format should follow the \texttt{pyvalem} program (\href{https://github.com/xnx/pyvalem}{https://github.com/xnx/pyvalem}).
    \item term: Directly from the term column in levels data. The trailing `*' should be removed if applicable. Format should be consistent with \texttt{pyvalem}.
    \item qn:parity: Terms of odd parity (those ending in * in the original data sources) are marked with \texttt{-} in this field; those of even parity are marked with \texttt{+}.
    \item Abbr: Only in Kurucz-based files; indicates whether the level is experimentally identified (\texttt{NI}) or purely calculated (\texttt{CA}).
\end{itemize}
\end{minipage}
\end{table}

%%%%%

\begin{table}
\caption{Extract from the \texttt{.states}  file for \textsc{NIST} database for Li I (neutral lithium atom)}
\centering
\begin{tabular}{crccccccc}
    \toprule
    $i$ & {$\tilde{E}$} & $g_{J}$  & $J$ & Unc &  qn:configuration & term & qn:parity \\
    \midrule
    1 & 0.000000    &  2  &   0.5  &   0.100000 &1s2.2s &  2S       &   + \\
    2 & 14903.660000    &  2 &    0.5 &    0.100000 & 1s2.2p    &2P  &   - \\
    3 & 14904.000000 &  4 &  1.5&  0.100000& 1s2.2p   &2P     &  - \\
    4 & 27206.120000 &  2   &  0.5 &    0.100000 &1s2.3s    & 2S   &    +\\
    5 & 30925.380000 & 2 &  0.5 &  0.100000 & 1s2.3p  & 2P  &    - \\
    \bottomrule
    
\end{tabular}
\label{tb20}
\end{table}

\begin{table}
\centering
\caption{Extract from the \texttt{.states} file for Kurucz database for Li I (neutral lithium atom)}
\label{tb9}
\begin{tabular}{ccrccccccc}
\toprule
$i$ & $\tilde{E}$ & $g_{J}$ & $J$ & Unc &$\tau$ & $g$factor & qn:Configuration & term & Abbr \\
\midrule
33 & 40439.020000 & 8  & 3.5 & 0.100000 & 4.0323e-07 & 0.889000 & 1s2.6g & 2G & NI \\
34 & 40439.070000 & 10 & 4.5 & 0.100000 & 6.0606e-07 & 0.909000 & 1s2.6h & 2H & NI \\
35 & 40439.070000 & 12 & 5.5 & 0.100000 & 6.0606e-07 & 1.091000 & 1s2.6h & 2H & NI \\
36 & 40967.990000 & 2  & 0.5 & 0.100000 & 2.7174e-07 & 2.002000 & 1s2.7s & 2S & NI \\
37 & 41217.580000 & 2  & 0.5 & 0.100000 & 7.6336e-07 & 0.666000 & 1s2.7p & 2P & NI \\
\bottomrule
\end{tabular}
\vspace{25pt}
\begin{minipage}{\textwidth}
\footnotesize
Abbr: Abbreviation inherited from Kurucz metadata. \texttt{CA} denotes calculated levels, and \texttt{NI} indicates experimentally identified levels; values are retained from the Kurucz dataset. 
\end{minipage}
\end{table}

\subsubsection{Trans Files}

Similar to the \texttt{.states} file format, the ExoMol \texttt{.trans} format consists of upper state ID $i$, lower state ID $f$, Einstein $A$ coefficient and Transition wavenumber $\tilde{\nu}_{if}$, seen Table \ref{tb4}. Tables \ref{tb21} and \ref{tb11} illustrates a representative sample (Li I) \texttt{.trans} file from \textsc{NIST} Database and Kurucz Database, respectively.  As discussed above, for some atoms the associated \texttt{.trans} files could not be produced due to the absence of the Einstein coefficient or lack of the relevant description.

\begin{table}
    \centering
    \caption{Specification of the \texttt{.trans} file \citep{jt939}.}
    \label{tb4}
    \begin{tabular}{cccc}
    \toprule
    \textbf{Field} & \textbf{Fortran format} & \textbf{C format} & \textbf{Description} \\
    \midrule
    $i$ & \texttt{I12} & \texttt{\%12d} & Upper state ID \\
    $f$ & \texttt{I12} & \texttt{\%12d} & Lower state ID \\
    $A$ & \texttt{ES10.4} & \texttt{\%10.4e} & Einstein $A$ coefficient in \(\text{s}^{-1}\) \\
    $\tilde{v}_{if}$ & \texttt{E15.6} & \texttt{\%15.6e} & Transition wavenumber in \(\text{cm}^{-1}\)
 \\
    \bottomrule
    \end{tabular}

 \vspace{10pt} % Adds spacing between the table and the explanatory text
\noindent\begin{minipage}{\textwidth}
\footnotesize
The \texttt{.trans} file contains 4 columns and is generated by combining data from the levels and lines data. Each valid row in the lines data corresponds to a row in the \texttt{.trans} file. The values for columns $i$ and $f$ are obtained from the levels data according to the following rules:
\begin{itemize}
\item{The ExoMol data standards use the given data formats for columns which should be separated by a single space.}
    \item $i$: Determined by matching $\tilde{E}_k$, $\text{conf}_k$, $J_k$, and $\text{term}_k$ in the lines data with $\tilde{E}$, Configuration, $J$, and Term in the levels data. The corresponding state ID in the levels data is assigned as the upper state ID. In cases where valid Term or $J$ values are not provided in the lines data but are available in the levels data, matching is performed only using $\tilde{E}_k$ and $\text{conf}_k$.
    \item $f$: Determined by matching $\tilde{E}_i$, $\text{conf}_i$, $J_i$, and $\text{term}_i$ in the lines data with $\tilde{E}$, Configuration, $J$, and Term in the levels data. The corresponding state ID in the levels data is assigned as the lower state ID. If valid Term or $J$ values are absent in the lines data but exist in the levels data, matching is performed using only $\tilde{E}_i$ and $\text{conf}_i$.
    \item $A$: Directly obtained from $A_{ki}$ (\(\mathrm{s}^{-1}\)) in the lines data. Rows without a valid $A_{ki}$ (\(\mathrm{s}^{-1}\)) are removed.
    \item $\tilde{v}_{if}$: Directly obtained from the $wn$ (\(\mathrm{cm}^{-1}\)) in the lines data.
\end{itemize}
\end{minipage}
\end{table}

\begin{table}
  \centering
  \caption{Extract from the transitions file for \textsc{NIST} database for Li I (neutral lithium atom), with columns formatted in ES15.6.}
  \label{tb21}
  \begin{tabular}{cccc}
    \toprule
    $i$ & $f$ & $A$     & $\tilde{v}_{if}$ \\
    \midrule
    15 & 13 & 6.9000E-02 & 6.800000E+00 \\ 
    14 & 13 & 4.6000E-03 & 6.800000E+00 \\ 
    14 & 12 & 6.4400E-02 & 6.800000E+00 \\ 
    21 & 19 & 6.5000E-01 & 9.600000E+00 \\ 
    21 & 20 & 4.6400E-02 & 9.600000E+00 \\ 
    \bottomrule
  \end{tabular}
\end{table}

\begin{table}
\centering
\caption{\label{tb11}Extract from the transitions file for Kurucz database for Li I (neutral lithium atom).}
\begin{tabular}{rrcc}
\toprule
$i$ & $f$ & $A$ & $\tilde{\nu}_{if}$ \\
\midrule
28  & 27  & 1.7100E+02 & 4.597000E+01 \\
28  & 26  & 3.4198E+01 & 4.597000E+01 \\
29  & 26  & 2.0512E+02 & 4.597400E+01 \\
109 & 111 & 2.5527E+03 & 4.745000E+01 \\
109 & 110 & 2.5527E+03 & 4.745000E+01 \\
94  & 95  & 3.9811E+03 & 6.193000E+01 \\
\bottomrule
\end{tabular}
\end{table}

\subsubsection{Partition functions}

The partition functions for both \textsc{NIST} and Kurucz databases are provided separately, for reasons discussed below, as \texttt{.pf} files in standard ExoMol format. For NIST,
the partition functions, $Q(T)$, were generated as a direct summation as given by:
\begin{equation}
Q(T) = \sum_{i} g_i e^{-\frac{c_2 \tilde{E}_i}{T}}
\label{e:pfsum}
\end{equation}
where $i$, in principle, runs over all states in the system; $\tilde{E}_i$ is the corresponding energy term value  in \(\mathrm{cm}^{-1}\), $c_2 = 1.438776877$ K/cm$^{-1}$ is the second radiation constant, and $T$ is the temperature in K. Finally,  $g_i = 2J_i + 1$ is the degeneracy of  state $i$, where we also follow the so-called astrophysicist's convention.  

It should be noted that there is an issue with the infinite number of bound levels in atoms that can lead to the partition function becoming infinite
under some circumstances, see the discussion given by \citet{22AlFe.partfunc}. Given the different number of states provided by the \textsc{NIST} and  Kurucz differ significantly, the two databases provide different approaches to this problem. %\red{sy: what is the solution?}

\textsc{NIST} provides partition functions at a given temperature computed as the direct sum over the tabulated energy levels. We mimicked this and tests showed that our partition functions agree well with   those provided by  \textsc{NIST}. Partition functions in \textsc{NIST} are therefore provided for each species on a 1 K grid up to 6000~K. Higher temperature partition functions can easily be computed using the program \textsc{PyExoCross} \citep{jt914}.

\begin{figure}
  \centering
  \includegraphics[width=0.31\linewidth]{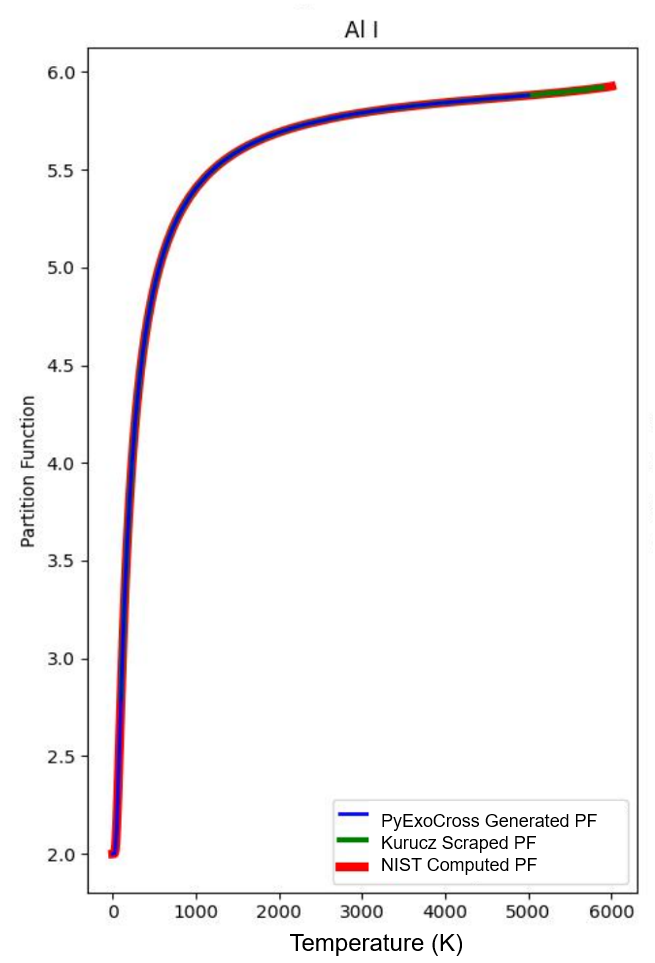}
  \includegraphics[width=0.31\linewidth]{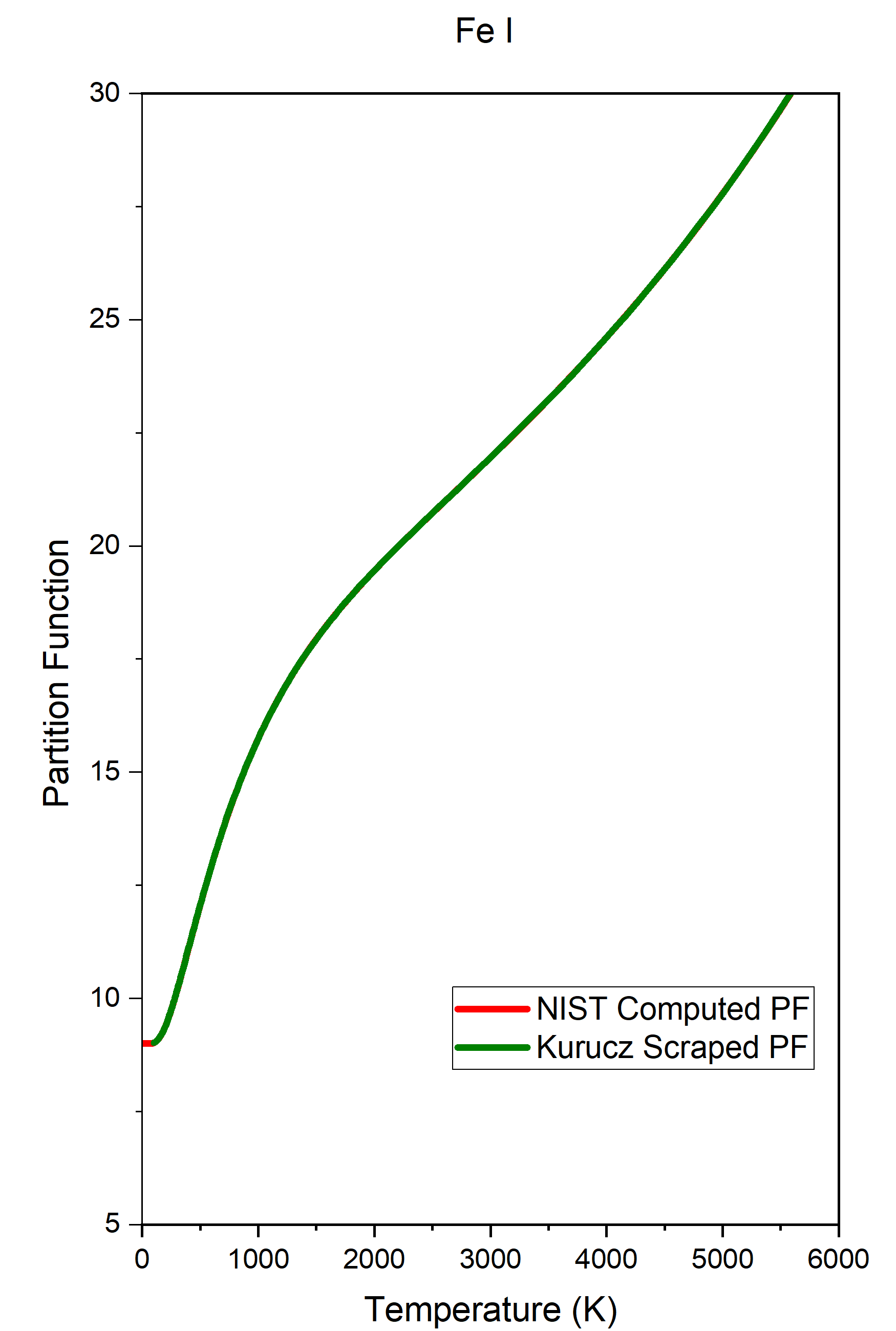}
  \includegraphics[width=0.29\linewidth]{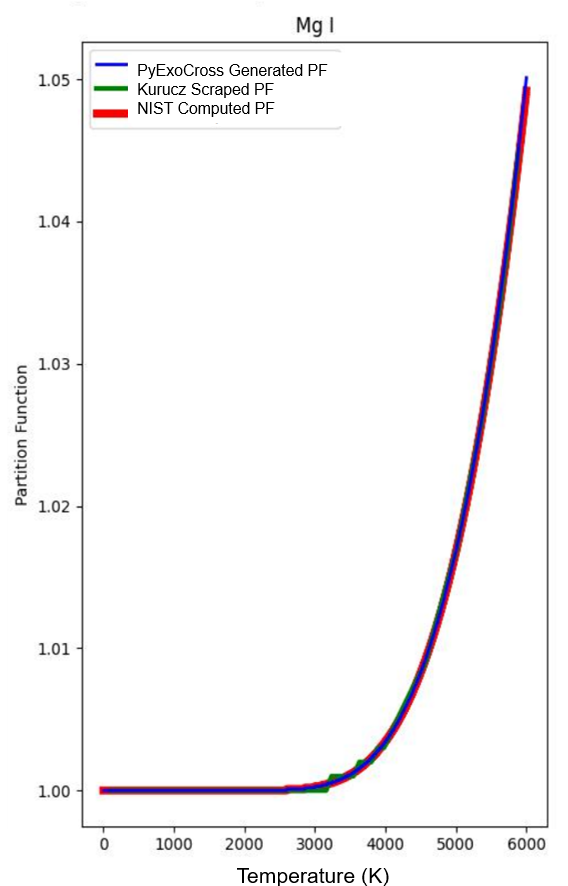}
  \caption{Comparison of partition functions for Al I, Fe I and Mg I, see text for further details.}
  \label{pyexoross}
\end{figure}

 %%\red{sy: Have I understood this correctly? Do we compute NIST Q(T) from scratch but keep the Kuruvz original values? Yes} 
%in a single file, \texttt{partfnxxyy.dat}, which contains multiple partition function values under different potential lowering conditions, expressed in \(\mathrm{cm}^{-1} / Z_{\mathrm{eff}}^2\). These conditions are represented by column headers with values of -500, -1000, -2000, -4000, -8000, -16000, and -32000. However, there is no clear documentation of this or criterion for selecting the appropriate columns. Partition functions from the first column were chosen and ExoMol-formatted to ExoMol format. 
In contrast, the Kurucz database explicitly provides partition functions, which we used directly. The temperature grid for the Kurucz partition functions  is coarser than that used for the \textsc{NIST} data but given the slow variation
in atomic partition functions should be sufficient for practical applications. Comparison between \textsc{NIST} and Kurucz values show them to be similar for simple systems but they can differ for systems with many low-lying excited states.  Figure \ref{pyexoross} compares \textsc{ExoAtom}  partition functions  generated for \textsc{NIST} by \textsc{PyExoCross} with those directly scraped from the Kurucz database. We could see that these three match very well. The agreement
between the representations of the partition function is good:
the maximum difference over the entire temperature range
considered is less than 0.2\% for Al, 0.1\% for Mg, and 0.1\% for Fe.

Table \ref{tb22} and Table \ref{tb13} illustrate representative samples of \texttt{.pf} files for Li I from the \textsc{NIST} and Kurucz databases, respectively. Both columns represent the partition function values at different temperatures. However, in the \textsc{NIST} database, the temperature steps are fixed with an interval of 1\,K, whereas in the Kurucz database, the temperature steps are non-uniform and vary across the dataset.

\begin{table}
\centering
\caption{\label{tb12} Specification of the ExoMol format \texttt{.pf} file (source from partfnxxyy.dat file in Kurucz).}
\begin{tabular}{llll}
\toprule
Field & Fortran Format & C Format & Description \\
\midrule
$T$ & F8.1 & \%8.1f & Temperature in K \\
$Q(T)$ & F15.4 & \%15.4f & Partition function \\
\bottomrule

\end{tabular}

 \vspace{10pt} % Adds spacing between the table and the explanatory text
\noindent\begin{minipage}{\textwidth}
\begin{itemize}
\item{The individual columns should be separated by an a additional single space.}
\end{itemize}
\end{minipage}
\end{table}

\begin{table}
    \centering
        \caption{Sample partition function calculation file from the \texttt{.states}  file for Ca I (neutral calcium atom) in the \textsc{NIST} database.}
    \begin{tabular}{cc}
    \toprule
    $T$ & $Q(T)$ \\
    \midrule
  5012.0   &       1.1737 \\
  5013.0    &      1.1739 \\
  5014.0      &    1.1741 \\
    5015.0       &   1.1743 \\
      5016.0    &      1.1744 \\
          5017.0    &      1.1746 \\
         
    \bottomrule
    \end{tabular}
    \label{tb22}
\end{table}

\begin{table}
\centering
\caption{Sample partition function calculation file from the \texttt{.states}  file for Ca I (neutral calcium atom) in the Kurucz database.}
\begin{tabular}{cc}
\toprule
$T$ & $Q(T)$ \\
\midrule
5012.0 & 1.1730 \\
5129.0 & 1.1950 \\
5248.0 & 1.2180 \\
5370.0 & 1.2450 \\
5495.0 & 1.2730 \\
5623.0 & 1.3050 \\
\bottomrule
\end{tabular}
    \label{tb13}
\end{table}

\section{Post Processing}

\textsc{PyExoCross} \citep{jt914} is a Python package that uses  ExoMol-formatted to generate spectra, cross-sections, lifetimes and opacities. With \textsc{PyExoCross}, spectra can be plotted based on the ExoMol-formatted atomic \texttt{.states} and \texttt{.trans} files to facilitate comparison with existing plots and other purposes.

\begin{figure}
  \centering
  \includegraphics[width=1\textwidth]{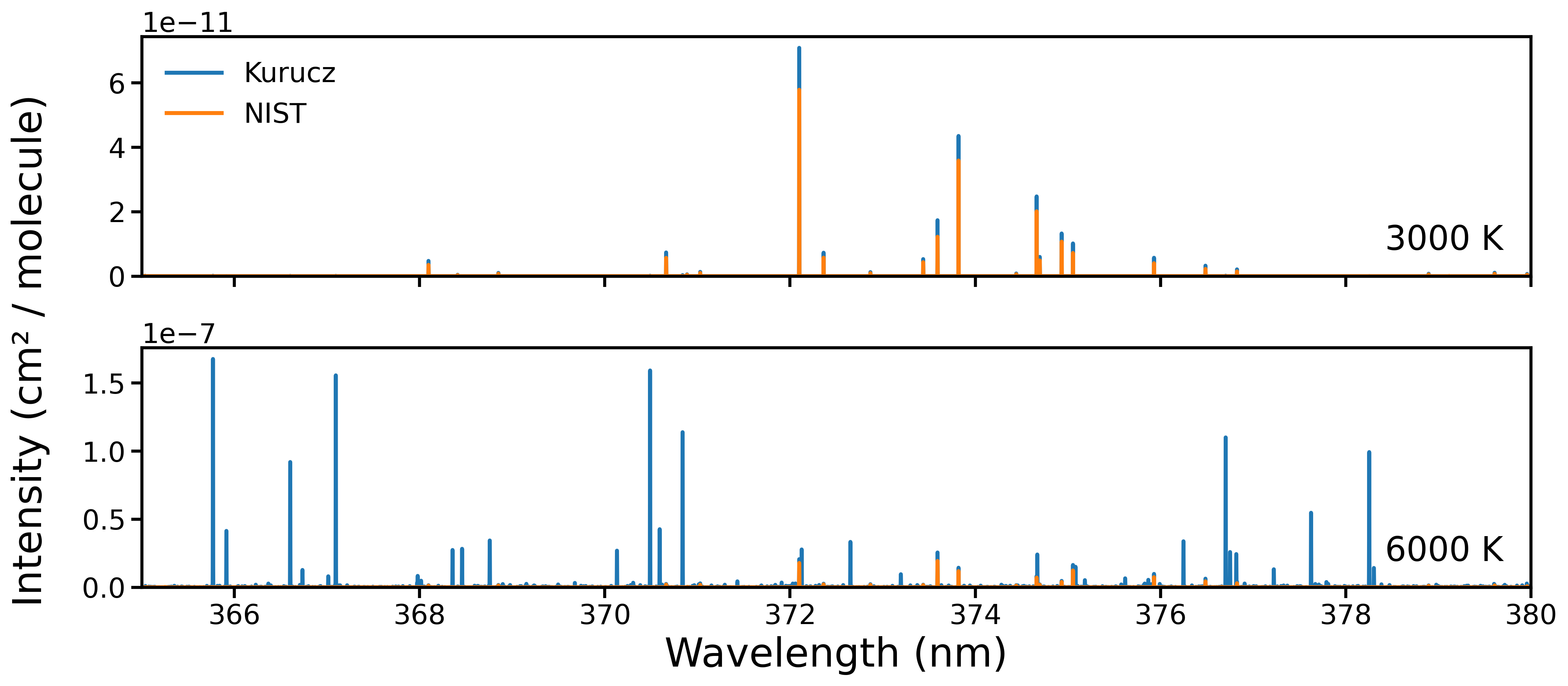}
  \caption{Comparison of \textsc{NIST} \citep{ralchenko2020development} and Kurucz \citep{Kurucz2017} Fe I emission spectra at 3000 K and 6000 K; the 6000 K spectrum contains many lines only present in the Kurucz dataset.}
  \label{Fe6000}
   \end{figure}

%, upper, middle panel, lower
The use of \textsc{PyExoCross} to generate spectra is illustrated in Figure~\ref{Fe6000},  which presents emission spectra at temperatures of 3000 and 6000~K generated from the \textsc{NIST} and Kurucz line lists within \textsc{ExoAtom}. 
The emission spectra were computed using pure Doppler line broadening with a bin size of 0.0003~nm, without any collisional or pressure broadening. At 3000 K,
both databases reproduce the key spectral features with good agreement for the peak position, while at 6000~K the more complete Kurucz Database shows many more lines.  There are also some differences for the predictied line intensities. For the \textsc{NIST} atomic database, uncertainties are provided for line positions (observed and Ritz wavelengths), but quantitative uncertainties for line intensities are generally not reported; however the \textsc{NIST} data for Fe I are substantially taken from \citet{5298TP} who in turn use the highly accurate measured oscillator strengths of
\citet{Blackwell}. Conversely the Kurucz data appears to be taken from \citet{Peterson_2014} who computed
oscillator strengths using the \citet{Cowan81} code. These sorts of differences are a feature of our comparisons
between the two sets of data; where lines exist in both codes, our assumption is the data provided by the
\textsc{NIST} atomic database should be favoured. At high 
temperatures Kurucz contains many strong transitions that are not recorded in NIST.

%\begin{figure}
%  \centering
%    \includegraphics[width=\textwidth]{Fe.png}
%    \includegraphics[width=\textwidth]{Fe_I.png}
%  \caption{Comparison of observed and simulated Fe I emission spectra. Upper panel: Emission spectra taken from \citet{Fe} with intensities in arbitrary units (a.u.); Lower panel: Emission spectra simulated with data from \textsc{NIST} and Kurucz \citep{ralchenko2020development,kurucz2011including,Kurucz2017}}
%    \label{figure_combined}
%   \end{figure}
 % %\red{sy: Is it possible to adjust the width, height, and the x-axis ranges in the ExoCross figure to match exactly the upper display? It is difficult to compare them. I can help by redrawing them in Origin, where overlaying is easy.}
  % %\red{I have no idea what global normalised means but please put NIST/Kurucz data on an absolute scale rather and a.u. It is the whole point of such data that the actual scale is known.}}

\section{Conclusions}
The \textsc{ExoAtom} database extends the \textsc{ExoMol} framework by providing high-accuracy atomic spectral data in a standardized format. This database integrates energy levels, radiative transitions, and partition functions for a wide range of neutral and ionized atoms, primarily sourced from the \textsc{NIST} and Kurucz databases. By structuring atomic data in the ExoMol format, \textsc{ExoAtom} enhances accessibility and ensures compatibility with established atomic datasets.

The current version of \textsc{ExoAtom} includes line lists for 79 neutral and 71  singly ionized elements from the \textsc{NIST} database, as well as 38 neutral and 37 singly ionized elements from the Kurucz database. The \textsc{NIST} database provides highly accurate data, whereas the Kurucz database offers a more complete dataset. Thus, \textsc{ExoAtom} does not recommend one dataset over another but instead presents both to accommodate different research needs. These datasets play a crucial role in modeling atomic processes in astrophysical environments, including stellar atmospheres, exoplanetary spectra, and the interstellar medium.
We note that while nearly all detections of atoms in exoplanets involve neutral or singly ionized species, some
studies have focused on more highly ionized atoms \citep{10LiYaFr}. Therefore, our plan is to expand the \textsc{ExoAtom}
database to include further ionization stages in due course. At present, the ExoAtom database has contains no line-broadening information. We note that the Kurucz database contains information on radiative, Stark and Van der Waals damping for most species,
concentrating largely on measured lines. We plan to add line-broadening parameters for
atoms and atomic ions in due course. 

The ExoMol group welcomes collaborations and contributions from external sources to further enrich their databases including the \textsc{ExoAtom} database. The  \textsc{ExoAtom} database \url{https://exomol.com/exoatom/} is publicly available through the ExoMol website.  We hope it will provide a valuable resource for spectroscopic modeling and interpreting observations.

%\red{sy: Please check that all figures and tables are referenced in the text and also that they appear in the order they are referenced.}

\section*{Acknowledgements}
We thank the attendees at the Royal Society Discussion meeting on the spectroscopy of exoplanets at high resolution
held at Sedgebrook Hall, Northamptonshire, UK in 2023 for suggesting to us that the ExoMol database should also
include atomic spectra. We thank Yuri Ralchenko of \textsc{NIST} and Robert Kurucz for help with the use of their datasets.
This work was supported by the European Research Council (ERC) under the European Union’s Horizon 2020 research and innovation programme via Advanced Grant 883830 and the STFC Project No. ST/Y001508/1. 

Since we started this project, Robert (Bob) Kurucz passed away on 1 March 2025 (see \citet{Dupree2025}) and the \textsc{NIST} Atomic Physics group has been closed. We wish to dedicate this paper to the talented atomic physicists whose work we make use of in this paper.

\section*{Data availability}

All the data discussed in this paper are freely available from the \textsc{ExoAtom} website \url{https://exomol.com/exoatom}.

\section{Conflict of interest}
Authors declare no conflict of interest.

\bibliographystyle{rasti}
\bibliography{journals_astro,jtj,other,rujiaref,tianyangref,atomic,linelists}
% if your bibtex file is called example.bib

\newpage

\appendix
\section{Example \texttt{adef.json} File  for the H Molecule}
\label{appendix:json_example}

\begin{lstlisting}[language=json]
{
    "species": {
        "atom": "H",
        "ordinary_formula": "H",
        "spectroscopic_notation": "H I",
        "charge": 0,
        "name": "hydrogen",
        "mass_in_Da": 1.00784
    },
    "isotope": {
        "iso_formula": "(1H)",
        "iso_name": "hydrogen",
        "mass": 1.00784,
        "spin": "1/2",
        "atomic mass number": 1
    },
    "dataset": {
        "name": "NIST",
        "version": 20240901,
        "doi": "",
        "max_temperature": 6000,
        "n_L_default": 0.5,
        "num_pressure_broadeners": 0,
        "nxsec_files": 0,
        "nkcoeff_files": 0,
        "dipole_available": false,
        "cooling_function_available": false,
        "specific_heat_available": false,
        "uncertainty_available": true
        "Ionisation": null,
        "states": {
            "number_of_states": 105,
            "max_energy": 109610.2232,
            "lifetime_available": false,
            "lande_g_available": false,
            "num_quanta": 1,
            "states_file_fields": [
                {
                    "name": "ID",
                    "desc": "Unique integer identifier for the energy level",
                    "ffmt": "I12",
                    "cfmt": "%12d"
                },
                {
                    "name": "E",
                    "desc": "State energy in cm^-1",
                    "ffmt": "F12.6",
                    "cfmt": "%12.6f"
                },
                {
                    "name": "gtot",
                    "desc": "State degeneracy",
                    "ffmt": "I6",
                    "cfmt": "%6d"
                },
                {
                    "name": "J",
                    "desc": "Total angular momentum quantum number, J (half-integer)",
                    "ffmt": "F7.1",
                    "cfmt": "%7.1f"
                },
                {
                    "name": "Unc",
                    "desc": "Uncertainty in the state energy in cm^-1",
                    "ffmt": "F12.6",
                    "cfmt": "%12.6f"
                },
                {
                    "name": "gfactor",
                    "desc": "Lande g-factor (optional)",
                    "ffmt": "F10.6",
                    "cfmt": "%10.6f"
                },
                {
                    "name": "qn:configuration",
                    "desc": "Configuration for the state",
                    "ffmt": "A12",
                    "cfmt": "%12s"
                },
                {
                    "name": "qn:LSCoupling",
                    "desc": "Term for the state",
                    "ffmt": "A8",
                    "cfmt": "%8s"
                },
                {
                    "name": "qn:parity",
                    "desc": "Parity for the state",
                    "ffmt": "A1",
                    "cfmt": "%1s"
                }
            ],
        },
        "transitions": {
            "number_of_transitions": 105,
            "number_of_transition_files": 1,
            "max_wavenumber": 109610.2,
            "transitions_file_fields": [
                {
                    "name": "i",
                    "desc": "Upper state ID",
                    "ffmt": "I12",
                    "cfmt": "%12d"
                },
                {
                    "name": "f",
                    "desc": "Lower state ID",
                    "ffmt": "I12",
                    "cfmt": "%12d"
                },
                {
                    "name": "A",
                    "desc": "Einstein A coefficient in s^-1",
                    "ffmt": "ES10.4",
                    "cfmt": "%10.4e"
                },
                {
                    "name": "Wavenumber",
                    "desc": "Transition wavenumber in cm^-1",
                    "ffmt": "E15.6",
                    "cfmt": "%15.6e"
                }
            ]
        },
        "partition_function": {
            "max_partition_function_temperature": 6000.0,
            "partition_function_step_size": 1,
            "fields": [
                {
                    "name": "T",
                    "desc": "Temperature in Kelvin",
                    "ffmt": "F8.1",
                    "cfmt": "%8.1d"
                },
                {
                    "name": "Q(T)",
                    "desc": "Partition function (dimensionless)",
                    "ffmt": "F15.4",
                    "cfmt": "%15.4d"
                }
            ]
        }
    }
}
\end{lstlisting}

\newpage
\section{Extract from the \textsc{ExoAtom} Master file showing the specifications for the H isotopes, Fe and Fe ion.}
\label{appendix:master}
\begin{lstlisting}[language=json]
{
  "ExoAtom": {
    "ID": "exoatom.all.json",
    "version": 20240930
  },
  "atoms": [
    {
      "name": "Hydrogen",
      "formula": "H",
      "num_isotopes": 3,
      "isotopes": [
        {
          "iso_slug": "1H",
          "iso_formula": "(1H)",
          "dataset": "NIST",
          "version": 20240601
        },
        {
          "iso_slug": "2H",
          "iso_formula": "(2H)",
          "dataset": "NIST",
          "version": 20240601
        },
        {
          "iso_slug": "3H",
          "iso_formula": "(3H)",
          "dataset": "NIST",
          "version": 20240601
        }
      ]
    },
    {
      "name": "Iron",
      "formula": "Fe",
      "num_isotopes": 1,
      "isotopes": [
        {
          "iso_slug": "56Fe",
          "iso_formula": "(56Fe)",
          "dataset": "NIST",
          "version": 20240601
        },
        {
          "iso_slug": "56Fe",
          "iso_formula": "(56Fe)",
          "dataset": "Kurucz",
          "version": 20240115
        }
      ]
    },
    {
      "name": "Iron Ion (I)",
      "formula": "Fe_p",
      "num_isotopes": 1,
      "isotopes": [
        {
          "iso_slug": "56Fe_p",
          "iso_formula": "(56Fe_p)",
          "dataset": "NIST",
          "version": 20240601
        },
        {
          "iso_slug": "56Fe_p",
          "iso_formula": "(56Fe_p)",
          "dataset": "Kurucz",
          "version": 20240115
        }
      ]
    }
    // ... other atoms ...
  ]
}
\end{lstlisting}


%%%%%%%%%%%%%%%%%%%%%%%%%%%%%%%%%%%%%%%%%%%%%%%%%%

%%%%%%%%%%%%%%%%% APPENDICES %%%%%%%%%%%%%%%%%%%%%



%%%%%%%%%%%%%%%%%%%%%%%%%%%%%%%%%%%%%%%%%%%%%%%%%%

% Don't change these lines
\bsp	% typesetting comment
\label{lastpage}
\end{document}